\documentstyle[aps,prl,floats,graphicx,epsfig]{revtex}
\begin{document}
\draft

\twocolumn[\hsize\textwidth\columnwidth\hsize\csname @twocolumnfalse\endcsname

%
\title{Field theory in superfluid $^3$He: \\
What are the lessons for particle physics, gravity, \\
and high-temperature superconductivity?}
\author{G.E. Volovik  }

\address{  Low Temperature Laboratory, Helsinki
University of Technology, Box 2200, FIN-02015 HUT, Espoo, Finland\\
and\\ Landau Institute for Theoretical Physics, Moscow, Russia}

\date{\today} \maketitle
\
\centerline{Communicated by Olli V. Lounasmaa}
\begin{abstract}

There are several classes of homogeneous Fermi-systems  which are
characterized by the topology
of the energy spectrum of fermionic quasiparticles: (1) Gapless systems
with a Fermi-surface;
(2) Systems with a gap in their spectrum; (3) Gapless systems with
topologically stable point
nodes (Fermi points); and (4) Gapless systems with topologically unstable
lines of nodes (Fermi
lines). Superfluid $^3$He-A and electroweak vacuum belong to the
universality Class (3). The
fermionic quasiparticles (particles) in this class are chiral: they are
left-handed or
right-handed. The collective bosonic modes of systems of Class (3) are the
effective gauge and
gravitational fields. The great advantage of superfluid
$^3$He-A is that we can perform experiments using this condensed matter and
thereby simulate
many phenomena in high energy physics, including axial anomaly,
baryoproduction, and
magnetogenesis. $^3$He-A textures induce a nontrivial effective metrics of
the space, where
the free quasiparticles move along geodesics. With $^3$He-A one can
simulate event horizons,
Hawking radiation, rotating vacuum, etc. High-temperature superconductors
are believed to belong
to Class (4). They have gapless fermionic quasiparticles with a
"relativistic" spectrum close to
gap nodes, which allows application of ideas developed for superfluid
$^3$He-A.
\end{abstract}
\
]

\section{Introduction}

It is now well understood that the Universe and its symmetry-broken ground
state
-- the physical vacuum -- may behave like a condensed matter system with
a complicated and possibly degenerate ground state
\cite{Hu,Wilczek,Jegerlehner,Jackiw}.

\begin{figure}[!!!t]
\begin{center}
\leavevmode
\epsfig{file=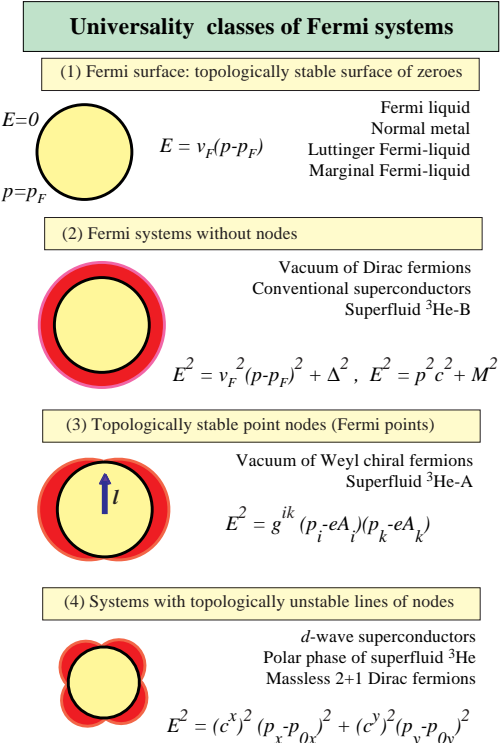,width=0.8\linewidth}
\caption[UniversalityClasses]
    {Universality classes of the fermionic ground state (vacuum).}
\label{UniversalityClasses}
\end{center}
\end{figure}

If the analogy of the quantum mechanical physical vacuum with condensed
matter systems is taken
seriously, the first question which arises is: Which system of condensed
matter
reproduces most closely the properties of the quantum vacuum? Since
particle physics
deals with interacting Fermi and Bose quantum fields, the system should be
fermionic. This
requirement excludes   superfluid $^4$He, which contains only Bose fields.
In Fermi systems,
such as metals, superconductors, and normal and superfluid $^3$He, in
addition to the fermionic
degrees of freedom which come from the bare fermions, electrons, and
$^3$He-atoms, respectively,
the quantum Bose fields appear as low-energy collective modes. Therefore,
these systems do
represent interacting Fermi and Bose quantum fields.

Which Fermi system is the best? To answer
this question we must first realize that the main feature which
differentiates between
various Fermi systems is the topology of the quasiparticle spectrum in the
low energy
(infra-red) corner. We will consider only systems whose ground state is
spatially
homogeneous -- this is one of the least disputed property of the physical
vacuum. When the
topology of the quasiparticle spectrum is taken into account, the
homogeneous Fermi systems can
be organized into very few classes (see Fig.~\ref{UniversalityClasses}).

\section{Systems with a Fermi surface}

\begin{figure}[!!!t]
\begin{center}
\leavevmode
\epsfig{file=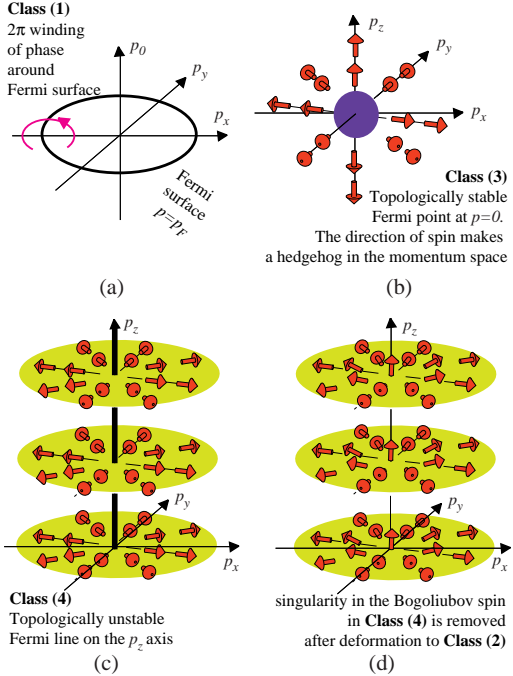,width=0.8\linewidth}
\caption[FermiSurfaceAsVortex]
    {(a) Winding of the propagator phase around the Fermi surface. For
simplicity the
$p_z$ coordinate is hidden so that the Fermi surface is the line
$(p_0=0, p=p_F)$ in the 2+1 momentum space. This line is a singularity,
which is similar to a vortex in a real 3D-space: The phase of the propagator
$G=(ip_0- (p_x^2 +p_y^2 -p_F^2)/2m)^{-1}$ changes by $2\pi$ around the line
in the
momentum space in the same manner as the phase of the order parameter
changes by $2\pi$ around
a vortex in the real space. (b) Fermi point at ${\bf p}=0$ in the 3D
momentum space
$(p_x,p_y,p_z)$. At this point the particle energy $E=cp$ is zero. A
right-handed particle is
considered with its spin  parallel to the momentum ${\bf p}$, {\it i.e.}
${\bf s}({\bf p})=
(1/2){\bf p}/p$. The spin makes a hedgehog in the momentum space, which is
topologically
stable.  (c) Fermi line -- topologically unstable manifold of zeroes -- is
shown in the 3D
momentum space
$(p_x,p_y,p_z)$. The (Bogoliubov) spin (arrows) is confined into the
$(p_x,p_y)$ plane and has
a singularity on the $p_z$ axis. (d) This singularity can be removed by a
continuous
transformation. The spin escapes into a third dimension ($p_z$) and becomes
well defined on the
$p_z$ axis. As a result, the quasiparticle spectrum becomes fully gapped
(the "relativistic"
fermion acquires the mass).}
\label{FermiSurfaceAsVortex}
\end{center}
\end{figure}

The most common universality class is made of fermionic systems which have
a Fermi surface
(FS). Any collection of weakly interacting fermionic particles belongs to
this class. In the extreme limit of a
noninteracting Fermi gas, with an energy spectrum
$E(p)=p^2/2m-\mu$,  where  $\mu$ is the chemical potential, the FS bounds
the volume
in the momentum space where  $E(p)<0$ and where the quasiparticle states
are all occupied at
$T=0$. In this isotropic model the FS is a sphere of radius
$p_F=\sqrt{2m\mu}$. It is remarkable that the FS survives even if
interactions between particles are introduced. This stability is a topological
property of the FS which is reflected in the Feynman quantum mechanical
propagator
$G =(z-{\cal H})^{-1}$ for the particle (the one-particle Green's function).

Let us write the propagator for a given momentum ${\bf p}$ and for the
imaginary frequency,
$z=ip_0$ (The imaginary frequency is introduced to avoid the conventional
singularity of the propagator at $z=E(p)$). For noninteracting particles
the propagator has the
form  $G=(ip_0 -E(p))^{-1}$. Obviously there is still a singularity:  On the
hypersurface
$(p_0=0, p=p_F)$ in the 4-dimensional space $(p_0, {\bf p})$ the propagator
is not well
defined. What is important is that this singularity is stable: The phase
$\Phi$ of the Green's
funtion $G=|G|e^{i\Phi}$ changes by
$2\pi$ around the path embracing this surface in the 4D-space (see Fig.
\ref{FermiSurfaceAsVortex}), and the phase winding number is robust towards
any perturbation.
Thus the singularity of the Green's funtion on the 2D-surface in the
momentum space is
preserved, even when interactions between particles are introduced.

\begin{figure}[!!!t]
\begin{center}
\leavevmode
\epsfig{file=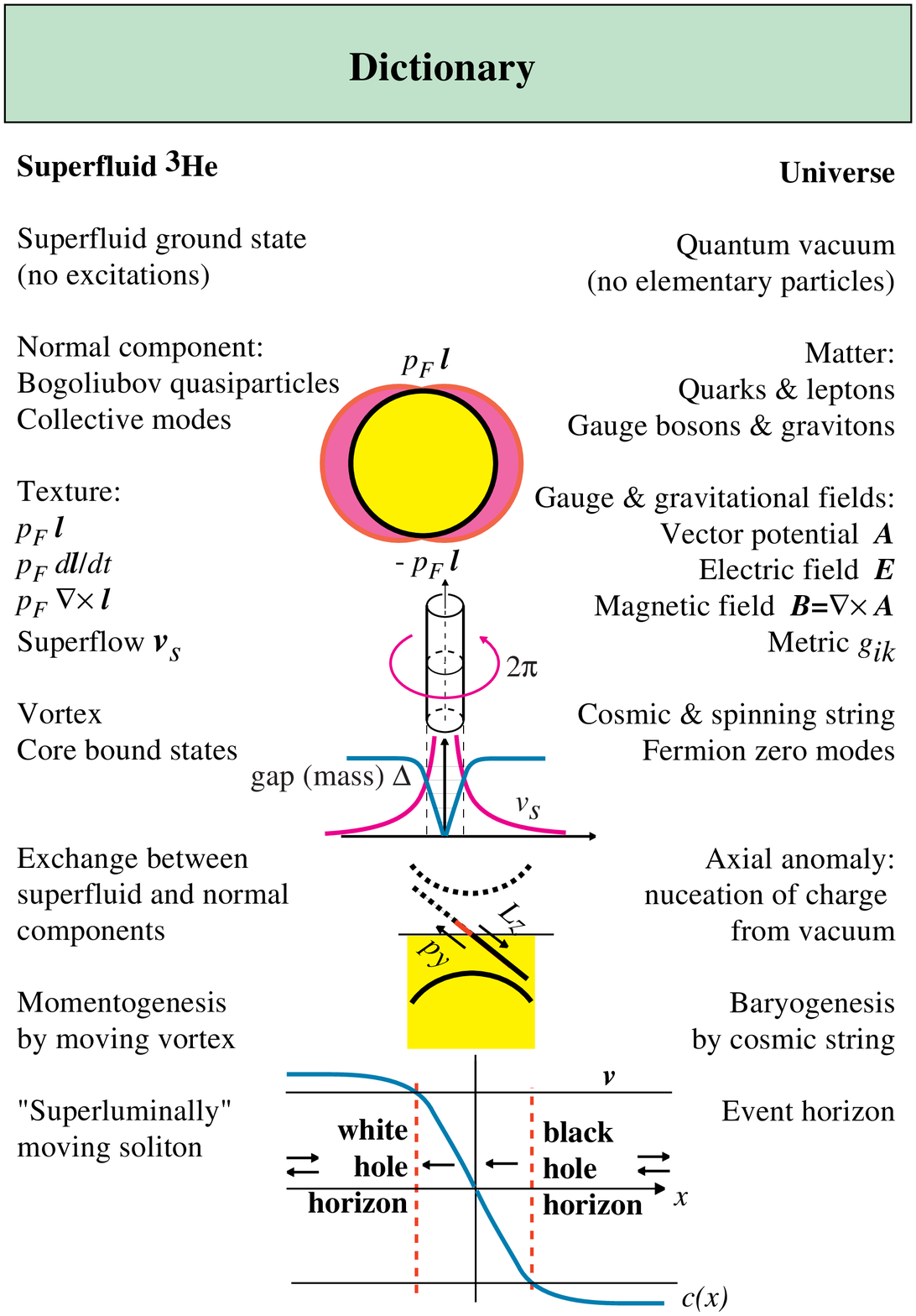,width=0.8\linewidth}
\caption[Dictionary]
    {Dictionary}
\label{Dictionary}
\end{center}
\end{figure}

Exactly the same topological conservation of the winding
number leads to the stability of the quantized vortex in superfluids and
superconductors, the
only difference being that, in the case of vortices, the phase winding
occurs in the real space
(see Fig. \ref{Dictionary}), instead of the momentum space. The connection
between the
topology in real space and the
topology in momentum space  is, in fact, even deeper (see {\it  e.g.}
Refs.\cite{VolovikMineev1982,Grinevich1988}).

\begin{figure}[!!!t]
\begin{center}
\leavevmode
\epsfig{file=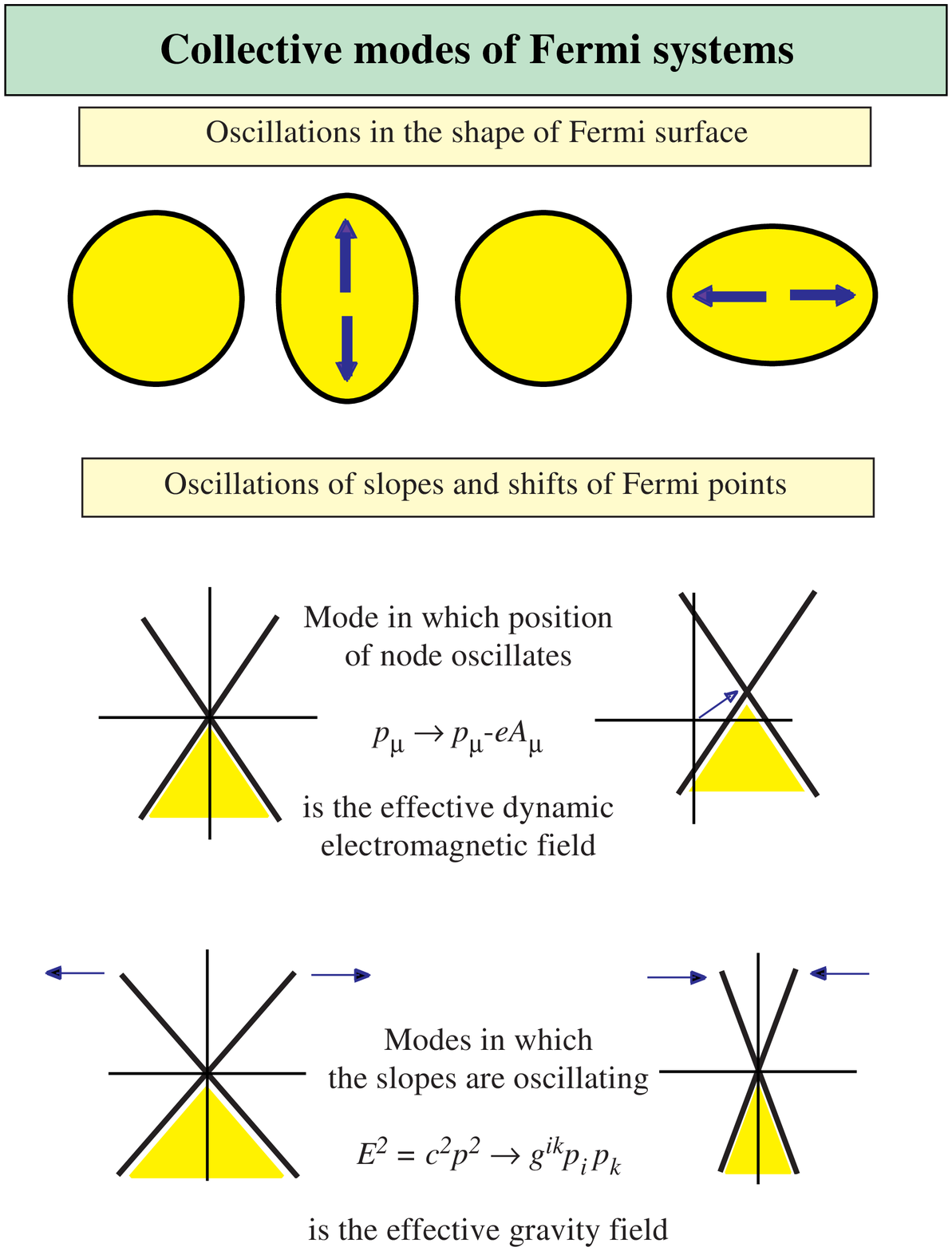,width=0.8\linewidth}
\caption[CollectiveModes]
    {Collective modes of fermionic systems}
\label{CollectiveModes}
\end{center}
\end{figure}

The topology of the propagator in the 4D-momentum space is thus essential
for the Landau
theory of an interacting Fermi liquid; it confirms the assumption that in
Fermi liquids
the spectrum of quasiparticles at low energy is similar to that of
particles in a Fermi gas.
The interactions do not change the topology of the fermionic spectrum  but
they produce the
effective field acting on a given particle by the other moving particles.
While this
effective field cannot destroy the FS owing to its topological stability,
it can
shift its position locally. Therefore, a collective motion of the particles
is seen
by an individual quasiparticle as a dynamical mode of the FS. These bosonic
oscillative modes
are known as different harmonics of the zero sound. An example is shown in
the upper part
of Fig. \ref{CollectiveModes}.

\section{Systems with a Fermi point}

\begin{figure}[!!!t]
\begin{center}
\leavevmode
\epsfig{file=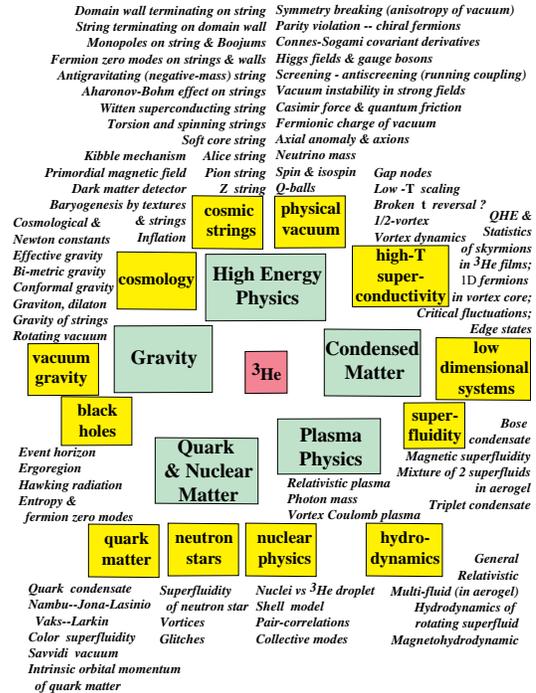,width=0.8\linewidth}
\caption[Connections]
    {Helium-3-centric Universe: Connections of  superfluid $^3$He to other
branches of physics.
For each item one can find the analogous phenomenon in superfluid phases of
$^3$He.
The analogies can result from the symmetry, from topology in real space or
from topology in
momentum space. In many cases they are described by the same equations.}
\label{Connections}
\end{center}
\end{figure}

Although the systems we have discussed contain fermionic and bosonic
quantum fields, this is
not the relativistic quantum field theory we need: There is no Lorentz
invariance
and the oscillations of the Fermi surface do not resemble the gauge field
even remotely.
The situation is somewhat better for Class (2), {\it i.e.} for fermionic
systems with fully gapped
spectra; examples, which provide useful analogies with Dirac fermions and
spontaneously
broken symmetry in quantum fields, are conventional superconductors
\cite{Nambu} and superfluid $^3$He-B \cite{ColorSuperfluidity}. The latter
also serves as a
model system for simulations of many phenomena in particle physics and
cosmology
(see Fig.~\ref{Connections}),  including experimental verification
\cite{MiniBigBang} of the
Kibble mechanism describing formation of cosmic strings in the early
Universe \cite{Kibble}.

However, we proceed to Class (3) which most fully exhibits the fundamental
properties needed for a realization of the relativistic quantum fields,
analogous to those in
particle physics and gravity.

Class (3) systems, whose representatives are superfluid
$^3$He-A and the vacuum of relativistic left-handed and right-handed chiral
fermions, is
characterized by points in the momentum space where the (quasi)particle
energy is zero. In
particle physics the energy spectrum $E({\bf p})=cp$ is characteristic of
the massless
neutrino (or any other chiral lepton or quark in the Standard Model of
electroweak interactions) with $c$ being the speed of light. The energy of
a neutrino is zero at
point ${\bf p}=0$ in the 3D momentum space. In condensed matter systems
such point
nodes have been realized first in superfluid $^3$He-A, which we discuss
later. The Hamiltonian
for the neutrino -- the massless spin-1/2 particle -- is a $2\times 2$
matrix $ {\cal H}=\pm
c\vec \sigma\cdot {\bf p}$, which is expressed in terms of the Pauli spin
matrices $\vec
\sigma$. The sign $+$ is for a right-handed particle  and $-$  for a
left-handed one:
the spin of the particle is oriented along or opposite to its momentum,
respectively.

Let us again consider the propagator of the particle
$G =(ip_0-{\cal H})^{-1}$ on the imaginary frequency axis, $z=ip_0$. One
can see that this
propagator still has a singularity, which is now not on the surface but at
point
$(p_0=0,{\bf p}=0)$ in the 4D momentum space. It is important that this
Fermi point
is not simply the  shrinked Fermi surface which is topologically unstable and
can disappear. The discussed points control the topological stability; they
cannot be
destroyed by external perturbations of the system.

Such stability can be visualized if one
considers the behavior of the particle spin ${\bf s}({\bf p})$ as a
function of its
momentum ${\bf p}$ in the 3D-space   ${\bf p}=(p_x,p_y.p_z)$. For
right-handed particles
${\bf s}({\bf p})={\bf p}/2p$, while for left-handed ones ${\bf s}({\bf
p})=-  {\bf
p}/2p$. In both cases the spin distribution in the momentum space looks
like a hedgehog (see
Fig. \ref{FermiSurfaceAsVortex}b), whose spines are represented by spins:
spines point outward
for the right-handed particle and inward for the left-handed one. In the
3D-space the hedgehog
is topologically stable: There exists an integer topological invariant,
which supports the
stability of the Fermi point in the same manner as the conservation of the
winding number is
responsible for the stability of a vortex line and the Fermi surface. This
invariant can be
expressed in terms of the propagator \cite{exotic}.

The consequence is the following: The effective fields acting on a given
particle
due to interactions with other moving particles cannot destroy the Fermi
point. They lead to
a shift in its position in the momentum space and to a change of the slopes of
the energy spectrum (see Fig.
\ref{CollectiveModes}).  This means that the low-frequency collective modes
in such Fermi
liquids are the propagating collective oscillations of the positions of the
Fermi point and of
the slopes at the Fermi point. The former is felt by the right- or the
left-handed
quasiparticles as the gauge (electromagnetic) field, because the main
effect of the
electromagnetic field
$A_\mu=(A_0,{\bf A})$ is just a dynamical change in the position of zero in
the energy
spectrum:
$(E-eA_0)^2=c^2 ({\bf p}-e{\bf A})^2$.

The latter, {\it i.e.} the change of the slope, corresponds to
a change in the speed of light, which can be different for different
directions in space:
$c^x=c+\delta c^x$,
$c^y=c+\delta c^y$,
$c^z=c+\delta c^z$. In a more general consideration the energy spectrum in
the perturbed state
is expressed in terms of the matrix of slopes, $E^2=g^{ik}p_ip_k$. In the
physical sense
this matrix is the metric tensor: The quasiparticles feel the inverse
tensor $g_{ik}$ as
the metric of the effective space in which they move along the geodesic
curves with the interval
$ds^2=-dt^2+g_{ik}dx^idx^k$. Therefore, the collective modes related to the
slopes play the part
of the gravity field (see Fig. \ref{CollectiveModes}).

The most general form of the energy spectrum
close to the Fermi point, {\it i.e.} at low energies, is $g^{\mu\nu}(p_\mu
-eA_\mu)(p_\nu
-eA_\nu)=0$, which describes a relativistic massless (actually chiral)
particle moving in the
electromagnetic and gravity fields. It is most important that this is the
general form of the
energy spectrum in the vicinity of any Fermi point, even if the underlying
Fermi system is not
Lorentz invariant; superfluid $^3$He-A is an example. The fermionic
spectrum necessarily
becomes Lorentz invariant near the Fermi point, {\it i.e.} this invariance
is not fundamental
but a low-energy property of any system with a Fermi point.

Another important property which results from the above equation is that
the fermionic
propagator near the Fermi point is gauge invariant and even invariant under
general
coordinate transformations (general covariance). For example, the local
phase transformation of
the wave function of the fermion, $\Psi \rightarrow \Psi e^{ie\alpha({\bf
r},t)}$ can be
compensated by a shift of the "electromagnetic" field $A_\mu \rightarrow
A_\mu + \partial_\mu
\alpha$. Such invariances are usually attributed  to fundamental properties of
electromagnetic ($A_\mu$) and gravitational ($g^{\mu\nu}$) fields, but here
they arise
spontaneously as low-energy phenomena.

What about equations for these collective bosonic modes, $A_\mu$ and
$g^{\mu\nu}$: Are they also gauge invariant, {\it i.e.} invariant under
transformation $A_\mu \rightarrow A_\mu + \partial_\mu
\alpha$ ? Also, do they obey the general covariance? In other words, do
they correspond to
Maxwell and Einstein equations for electromagnetic and gravitational
fields, respectively?
The answer to this question depends on the structure of the underlying
Fermi system. Let us
discuss this.

The effective Lagrangian for the collective modes is obtained by
integrating over the
vacuum fluctuations of the fermionic field in the presence of the
collective bosonic fields.
This principle was used by Sakharov and Zeldovich to obtain an effective
gravity
\cite{Sakharov} and an effective electrodynamics \cite{Zeldovich}, both
arising from
fluctuations of the fermionic vacuum.

Let us suppose that the main contribution to the effective action comes
from the vacuum fermions
whose momenta ${\bf p}$ are concentrated near the Fermi point. Since these
``relativistic''
fermions, moving in "gauge" and "gravity" fields, obey gauge invariance and
general
covariance, the result of the integration -- the effective Lagrangian for
the bosonic fields --
is also Lorentz invariant, gauge invariant, and even has a covariant form.
In this case the
obtained effective Lagrangian does give Maxwell equations for $A_\mu$
\cite{Zeldovich} and
Einstein equations for $g_{\mu\nu}$ \cite{Sakharov}, so that the
propagating bosonic
collective modes do represent the gauge bosons and  gravitons.

In the extreme limit,  when the massless relativistic fermions are
overdominating in vacuum,
a new invariance is revealed -- the conformal inariance: The conformal
transformation
$g_{\mu\nu}\rightarrow ag_{\mu\nu}$ leaves the massless fermions intact, as
a result the
effective action for gravity becomes the conformly invariant Weyl action.
Weyl gravity is a
viable rival to Einstein gravity  in modern cosmology \cite{Mannheim,Edery}.

Thus, if two
requirements are fulfilled -- (i) the fermionic system has a Fermi point
and (ii) the main
physics is concentrated near this Fermi point  -- the system acquires at
low energy all the
properties of modern quantum field theory: chiral fermions, quantum gauge
fields, and
gravity. All these ingredients are actually low-energy (infra-red) phenomena.

As a practical realization of Class (3) Fermi systems in condensed matter,
let us
consider excitations in $^3$He-A (see Fig. \ref{Dictionary}). After the
transition
to the superfluid state the Fermi surface disappears, a gap appears instead
in the quasiparticle
energy spectrum. Distinct from conventional superconductors belonging to
Class (2), the gap
has nodes at the north and south poles of the former Fermi surface (at ${\bf
p}=\pm p_F{\hat{\bf l}}$, where ${\hat{\bf l}}$ is the direction of
spontaneous angular
momentum in $^3$He-A). Each node is a topologically stable Fermi point with
left-handed
quasiparticles near the north pole and right-handed quasiparticles near the
south pole. Another
example of the Fermi point in condensed matter has been discussed for
gapless semiconductors
\cite{Abrikosov}.

Close to the gap nodes, {\it i.e.} at energies $E\ll\Delta_0$, where
$\Delta_0$ is the maximal
value of the gap in $^3$He-A, playing the part of
the Planck energy, the quasiparticles obey the relativistic equation
$g^{\mu\nu}(p_\mu - e A_\mu)(p_\nu - e A_\nu)=0$. Here $e=\pm$ is the "electric charge" and
simultaneously the chirality of the quasiparticles. The effective
electromagnetic field is
induced by the dynamical ${\hat{\bf l}}$-field and by the velocity ${\bf
v}_s$ of
the superfluid quantum vacuum of $^3$He-A: ${\bf A}=p_F{\hat{\bf l}}$,
$A_0=p_F {\bf v}_s\cdot
{\hat{\bf l}}$. This means, {\it  e.g.} that the texture of the ${\hat{\bf
l}}$-vector is felt by
quasiparticles as the magnetic field according to equation ${\bf B}=p_F
\vec\nabla\times{\hat{\bf l}}$ (see Fig. \ref{Dictionary}). Moreover, in
the low-energy limit,
the $A_\mu$ field does obey Maxwell equations: The integration over the
vacuum fermions is
concentrated near the Fermi point due to logarithmic divergence, known in
particle physics as the zero charge effect.

The metric of the effective space, where the chiral quasiparticles move
along the geodesic
curves, has the components:
$g^{ik}= c_\perp^2 (\delta^{ik} - \hat l^i \hat l^k) +  c_\parallel^2 \hat l^i
\hat l^k -v_s^iv_s^k$, $g^{00}=-1$, $g^{0i}=v_s^i$. The quantities
$c_\parallel$ and
$c_\perp$ -- the velocities of "light" propagating
along and transverse to  ${\hat{\bf l}}$ -- are expressed by the "fundamental"
parameters of $^3$He-A: $c_\parallel= p_F/m^*$,  $c_\perp= \Delta_0/p_F$,
where
$m^*$ and $p_F$ are the mass and Fermi momentum, respectively, of
quasiparticles in the normal
$^3$He. Since $\Delta_0 \ll   E_F\sim p_F^2/2m^*$,  the "Einstein" and "Weyl"
actions for $g_{\mu\nu}$ are highly contaminated by many noncovariant
terms, which come from the
integration over the "nonrelativistic" high energy degrees of freedom in
the region $\Delta_0
<E<E_F$.

In spite of the absence of general covariance in general, many different
properties of the
physical vacuum with a Fermi point, whose direct observation is still far
from realization,
can be simulated in $^3$He-A. One of them is the chiral anomaly, which
allows the nucleation of
the fermionic charge from the vacuum
\cite{Adler1969,BellJackiw1969} (see Figs.
\ref{Dictionary},~\ref{Connections}). Since the
chiral anomaly is a low-energy phenomenon,
the anomaly equation in $^3$He-A has gauge invariant and general covariant
form, and thus
exactly coincides with that derived by  Adler and by Bell and Jackiw
\cite{Adler1969,BellJackiw1969}. This equation has been verified in several
$^3$He experiments
\cite{BevanNature,AxialAnomaly}. In particle physics, the only evidence of
axial anomaly is
related to the decay of the neutral pion $\pi^0\rightarrow 2\gamma$,
although the anomaly is
much used in different cosmological scenaria explaining an excess of matter
over antimatter in
the Universe (see review \cite{Trodden}).

The advantage of $^3$He-A for such
simulations is that  the theory of this superfluid is in some sense
complete: At least in
principle one can derive everything from the bare $^3$He atoms interacting
via a
known potential. This is why there is no cut-off problem: we know (or can
calculate from
first principles) what happens not only in the low energy limit, where the
fermionic spectrum is
gauge invariant or covariant, but also at higher energy, where the Lorentz
and gauge
invariances are violated. This allows us to investigate problems which require
knowledge of physics beyond the Planck cut-off, {\it  e.g.} the quantum
effects related to the event
horizon of black holes. It is also important that there is a variety of
textures in superfluid
$^3$He-A which allow us to simulate the event horizon and ergoregion when
the texture moves with
velocity exceeding the local speed of "light" \cite{JacobsonVolovik} or rotates
\cite{Calogeracos} (see Figs. \ref{Dictionary},~\ref{Connections}).

\section{Systems with a Fermi line}

The high-temperature superconductors in cuprates most probably
contain zeroes in their quasiparticle energy spectrum. The ARPES experiments
\cite{Ding} show that these are lines in the 3D momentum
space where the quasiparticle energy is zero  or, equivalently,  point
zeroes in
the 2D CuO$_2$ planes.  The high-T superconductors thus belong to Class (4) of
systems with Fermi lines: the dimension $D$ of the manifold of zeroes is
$1$, which is
intermediate between a Fermi surface with
$D=2$ and a Fermi point with $D=0$.

As in the other two classes, all  low-energy (or infra-red)
properties of cuprate superconductors are determined by zeroes. In particular,
the density of the fermionic states with energy $E$ is determined by the
dimension of the zeroes: $N(E)=\sum_{\bf p}\delta (E-E({\bf p})) \sim
E^{2-D}$.
Many low-temperature properties of these superconductors are obtained from
a simple scaling,
known in $^3$He-A (this represents one of the connections in
Fig.~\ref{Connections}). For
example, an external magnetic field $B$ has the dimension of $E^2$. At
finite $B$, the density
of states is nonzero even at $E=0$ and equals
$N(0,B)\sim B^{(2-D)/ 2}$ \cite{FermionsOnQuantized}. An experimental
indication of such scaling
with
$D=1$ was reported for YBa$_2$Cu$_3$O$_7$ in Ref.~\cite{Revaz}.

The energy spectrum of quasiparticles near each of the 4 gap nodes in
Fig.~\ref{UniversalityClasses} can be written as
$E=\tau_1 c^x(p_x-eA_x) + \tau_3 c^y(p_y-eA_y)$, where $\vec\tau$'s are the
Pauli matrices in
the Bogoliubov-Nambu particle-hole space. The "speeds" of light $c^x$ and
$c^y$ are the
"fundamental" characteristics determined by the microscopic physics of
cuprates; ${\bf
A}$ is the effective (not electromagnetic) field, which indicates the
position of the nodal
lines in the momentum space. This means that the system belongs to the same
class as the 2+1
quantum electrodynamics with massless fermions.

The lines of zeroes generally have no topological stability: The singular
line in the momentum
space from which the spines (now the vector  $\vec\tau$) point outward (see
Fig.
\ref{FermiSurfaceAsVortex}c) can be elmininated by the escape of the
$\vec\tau$-vector to a
third dimension. This may be accomplished by an operation similar to the
folding of an
umbrella  (see Fig.
\ref{FermiSurfaceAsVortex}d).

Existence of the nodal lines can be prescribed, however, by the symmetry of
the ground
state. There are many  nontrivial classes  of
superconductors, whose symmetry supports  the existence of nodal lines in
symmetric
positions in the momentum space \cite{VolovikGorkov}. The symmetry
violating  perturbations,
such as impurities, an external magnetic field, {\it etc}., destroy the
lines of zeroes
\cite{Grinevich1988}. One could expect different types of transformations
of these lines of
zeroes which depend on the perturbation. Impurities, for example, can : (i)
produce the gap in
the fermionic spectrum \cite{Pokrovsky1995}, thus transforming the system
to Class (2) (see
Fig.
\ref{FermiSurfaceAsVortex}d);
(ii) lead to a finite density of states \cite{Sun1995}, thus transforming
the system to
Class (1); (iii) produce zeroes of fractional dimension, which means that
the exponent
in the density of states $N(E)\propto
E^{2-D}$ is non-integral  \cite{Nersesyan1994} and thus corresponds to a
fractional
dimension $D$ of the manifold of zeroes; and (iv) lead to localization
\cite{Lee1993}. An open
question is: Can the quantum fluctuations do the same, in particular, can
they change the
effective dimension of the zeroes?

\section{Conclusion}

The fermionic systems with topologically stable Fermi points have a
remarkable property:
In the low-energy corner the system exhibits an enhanced symmetry. The
Lorentz invariance,
general covariance, gauge invariance, and conformal invariance all appear
spontaneously in this
corner and  bring with them chiral relativistic fermions, gauge fields, and
gravity.  All are
low-energy phenomena, which are absent at higher energies. In particular,
this suggests that
gravity exists only in the infrared limit, {\it i.e.} only low-energy
gravitons can be
quantized \cite{Hu96}.

Distinct from the string theory,  which also gives rise to gravitation, the
Fermi point
mechanism does not require a high dimensionality for the space-time:  The
topologically stable
Fermi point is a property of the conventional 3+1 dimensional space-time.

There are actually two main guesses about the symmetry at high energies:
(i) Conventional
wisdom prescribes a higher symmetry at higher energies: $SO(10)$,
supersymmetry, etc.; (ii) A
contrary conjecture is that all symmetries known in the Universe disappear
at higher
energies when the Planck energy is approached. This includes the Lorentz
invariance
\cite{Chadha}, whose violation at high energies can be the origin of the
observed neutrino
oscillations
\cite{Glashow}. The condensed matter analogy with Fermi points supports the
second guess.

At "very low" energies of the electroweak scale $E_{\rm ew} \sim 100$
GeV, the chiral fermions acquire masses and become the Dirac fermions of
Class (2).
There are also two main guesses how this happens and both can be described
in terms of the
Fermi points:  (i) The Standard Model of symmetry breaking.  From the
point of view of
momentum-space topology, it corresponds to the coalescence and mutual
annihilation of Fermi
points with opposite topological invariants.  (ii) In the alternative theory,
the mass matrix of fermions appears in the same way as the gauge field
(see {\it e.g.}
\cite{Martin,Sogami}).  Using the language of
the Fermi points,   the gauge fields, the Higgs fields, and Yukawa
interactions, all are
realized as shifts of positions of Fermi points corresponding to different
quarks and leptons.
It is interesting that this way of unification of gauge and Higgs fields,
which has been called
"generalized covariant derivative", is known in
$^3$He-A, where most of effective gauge fields come from the collective
modes of the order
parameter, {\it i.e.} from the Higgs field.

However, at the moment it is not clear whether the Fermi point is really
the true way of how all
the low-energy phenomena arise in the physical vacuum. Before making a
conclusion one should manage to construct a scenario of how the 45 chiral
fermions of 3
generations (or 48 fermions, if neutrinos have a mass) and 12 gauge bosons
of the
$SU(3)\times SU(2) \times U(1)$ group of strong and electroweak
interactions can arise as
effective Fermi and Bose fields from the Fermi points.

$^3$He-A gives some hints that this may be possible. The Fermi point (say,
at the north pole) is
actually doubly degenerate owing to the ordinary spin of the $^3$He atom.
The double degeneracy
results in the $SU(2)$ effective gauge field, acting on quasiparticles near
the Fermi point
\cite{VolovikVachaspati,exotic}. This means that the higher symmetry groups
could be a
consequence of the Fermi point  degeneracy. For example, the 4-fold
degeneracy could
result in the 4 left-handed + 4 right-handed  fermionic species and
simultaneously in the
$SU(4)$ gauge group.

In this example, however, the number of bosons exceeds the number of
fermions. To obtain the correct number of bosons and fermions would
probably require a composite model for quarks and leptons.  There is,
however, another
possibility how to reduce the number of the effective gauge fields. Chadha
and Nielsen
\cite{Chadha} considered the massless electrodynamics with different
metrics for the left-handed
and right-handed fermions; their model thus violates the Lorentz
invariance. They found
that the two metrics converge to a single one as the energy is lowered.
Thus in the low-energy
corner the Lorentz invariance becomes better and better, and at the same
time the number of
independent bosonic modes decreases.

The violation of all invariances at high energy imposes another problem to
be solved: Why
are the corrections due to noninvariance extremely small at low energies?
Actually none
of the corrections have been experimentally identified so far. This means
that the
integration over the fermionic vacuum, which produces the action for the
gauge and gravity
fields, is very effectively concentrated near the Fermi points, where all
the symmetries are
present. There should be a special meachanism, such as an enhanced
quasiparticle  relaxation at
higher energy, which effectively switches off the nonsymmetric high-energy
contribution. The
same mechanism could be responsible for the absence of the cosmological
term in theEinstein
equations.

As we have already mentioned, with given physical parameters $^3$He-A is
not a good model
for such effective cancellation: While the Maxwell action is really
dominating at low energies
due to the logarithmically running coupling constant, the
Einstein action is polluted by the noncovariant terms, since the
contribution of the vacuum
fermions far from the Fermi point becomes dominating. To remove the
polluting terms, the integration must be spontaneously cut-off at energies
much below the
"Planck" scale, $E\ll \Delta_0$. This produces strong limitations on the
parameters of the
underlying condensed matter.

Nevertheless many phenomena related to the physical vacuum have been or
could be visualized in
$^3$He-A. There are many other connections between superfluid $^3$He and
the rest of physics
which should be exploited as well (see Fig.~\ref{Connections}).

\vfill\eject

\end{document}